\documentstyle[12pt]{article}
\begin{document}
{\raggedleft Bonn-TH-9523\\}
\bigskip
\title{\bf\huge
Exact diagonalization of the quantum supersymmetric $SU_q(n|m)$ model
}

\author{
{\bf Rui-Hong Yue$^a$\footnote{email address: yue@avzw02.physik.uni-bonn.de}
,Heng Fan$^{b,c}$,Bo-yu Hou$^c$}\\
\normalsize $^a$ Physikalisches Institut der Universit\"at Bonn\\
\normalsize Nussallee 12, 53115 Bonn, Germany\\
\normalsize $^b$ CCAST(World Laboratory)\\
\normalsize P.O.Box 8730,Beijing 100080,China\\
\normalsize $^c$ Institute of Modern Physics,Northwest University\\
\normalsize P.O.Box 105,Xian,710069,China
}

\maketitle

\begin{abstract}
We use the algebraic nested Bethe ansatz to solve the eigenvalue
and eigenvector problem of the supersymmetric $SU_q(n|m)$ model with
open boundary conditions. Under an additional condition that model is related to 
a multicomponent supersymmetric t-J
model. We also prove that the transfer matrix with open boundary
condition is $SU_q(n|m)$ invariant.
\end{abstract}
\maketitle
PACS: 7510J, 0520,0530


\newpage
\section{Introduction}
The integrability of  two-dimensional 
lattice models with periodic boundary condition is a consequence of the Yang-Baxter 
equation[1,2],  

\begin{eqnarray}
R_{12}(u-v)R_{13}(u)R_{23}(v)=R_{23}(v)R_{13}(u)R_{12}(u-v)
\end{eqnarray} 

\noindent where the R-matrix is the Boltzmann weight of the  
two-dimensional vertex model. As usual, $R_{12}(u)$,
$R_{13}(u)$
and $R_{23}(u)$ act in $C^n\otimes C^n\otimes C^n$ with 
$R_{12}(u)=R(u)\otimes 1$,$R_{23}(u)=1\otimes R(u)$,etc.

During the last years, much more attention has been paid on the investigation of integrable
systems with nontrivial boundary conditions, which was initiated by 
Cherednik[3] and Sklyanin[4]. They have 
introduced a systematic approach to handle the boundary problem in  which
the reflection equations appear. In   addition with the Yang-Baxter equation,
the reflection equations  ensure the integrability of open models. 
Using this approach, Sklyanin [4],  Destri and de Vega [5] solved  the spin- 
$1\over 2$ XXZ model with general boundary conditions by generalizing quantum 
 inverse scattering method. Under a particular choice of boundary conditions,
the Hamiltonian  is $U_q[sl(2)]$ invariant [6]. In [4],
Sklyanin assumed that the R-matrix is $P-$ and $T-$symmetric. Furthermore, 
the R-matrix satisfies unitarity and cross-unitarity properties. Because
only few models satisfy these properties, Mezincescu and Nepomechie [7]
extended Sklyanin's formalism to the $PT$-invariant systems. Thus, all  
trigonometric R matrices listed by Bazhanov [8] and Jimbo [9] can be related  
to 1-dimensional quantum spin chains in this formulism. Using the  unitarity and 
cross-unitarity properties of  Belavin's $Z_n$ elliptic
 R-matrix, we have constructed the open boundary  transfer matrix 
 with  one parameter [10,11].

On the other hand,  the study of open boundary conditions in 2-dimentional field
theory is related to the Sine-Gorden, Affine Toda and O(N) Sigma models 
[12,13,14,15]. Sklyanin generalized the hamiltonian to the case nonlinear partial
differential equations with local boundary conditions [15]. The reflection matrix 
is consistent with the integrability of the systems.

Recently, Foerster and Karowski have used the nested Bethe ansatz method
to find the eigenvalues and eigenvectors of the 
supersymmetric t-J model with open boundary conditions  and proved
its  $spl_q(2,1)$  invariance [16]. Gonzalez-Ruiz also solved this problem
with the general diagonal solutions of the reflection equation [17]. The investigated
model is a graded 15-vertex model characterized by two bosons and
one fermion. De Vega and Gonzalez-Ruiz
have also generalized the nested Bethe ansatz to the case of 
$SU_q(n)$-invariant chains[18].

The  graded veretx model was first proposed by Perk and Schultz [19]. In this model  all
variables take $m+n$ different values and the weights favor ferroelectric or
antiferroelectric configurations. In the references [20] and [21], the Bethe Ansatz
equations and  the exact free energy and excited expectrum of this model with periodic
boundary condition are found. The finit size 
correction shows the central charge of the model being $m+n-1$ (replacing $n$ in ref [20] by
$m+n-1$ ). In fact, the supersymmetric t-J model is a special Perk-Schultz model ($m=2,n=1$).
Under an appropiate boundary condition, the model enjoys beautiful structure as quantum 
group symmetry. This motives us to consider the general graded vertex model with
open boundary condition.

In this paper, we  use the nested Bethe ansatz method to find the 
eigenvalues and eigenvectors of the transfer matrix for a graded vertex 
model with open boundary conditions. The transfer matrix with fixed
boundary conditions is  proved to be $SU_q(n|m)$ invariant. 
When $m=1$, the model reduces into the  q-deformed version of the  
generalized supersymmetric t-J model with $n$ components. The
hamiltonian contains a spin hopping term, the nearest neighbour spin-spin interaction
and the contribution of boundary magnetic fields (see equation (24) ).
Now, we outline the contents of this paper.  In sect.2 we introduce the $SU_q(n|m)$
vertex model. We find the matrices $K^{\pm }$ which define boundary 
conditions and nontrivial boundary terms in the hamiltonian. 
The relation between the transfer matrix and the hamiltonian of the generalized 
supersymmetric t-J model is also discussed as an example. 
Sect.3 covers  to the diagonalization and the energy spectrum of the 
model with open boundary conditions in frame work of the nested Bethe Ansatz.
In sect.4 we show that the vertex model 
is a realization of the quantum supergroup $SU_q(n|m)$. A proof that 
the transfer matrix with open boundary conditions is $SU_q(n|m)$ 
invariant is given. In sect.5 the summary of our main results is presented
and some further problems are discussed. The appendix contains some detailed
calculation.

\section{The vertex model and integrable open boundary conditions}

Our starting point is a graded vertex model which  was introduced by Perk
and Schultz [19]. The thermodynamics of the model with periodic boundary condition
was studied in [20,21]. Some interesting application of this model in
quantum field theory was considered by Babelon, de Vega and Viallet [22].

The model is defined by vertex weights $R(u)$, whose  non-zero elements  are

\begin{eqnarray}
R^{aa}_{aa}(u)&=&sin(\eta +(-1)^{\epsilon _a}u),\nonumber \\
R^{ab}_{ab}(u)&=&sin(u)(-1)^{\epsilon_a\epsilon_b},a\not= b\nonumber \\
R^{ab}_{ba}(u)&=&sin({\eta })e^{iusign(a-b)}, a\not= b.\nonumber \\
\end{eqnarray}

\noindent where
\begin{eqnarray}
\epsilon _a=\left\{ 
\begin{array}{ll}
0, &a=1,\cdots ,m\\
1, &a=m+1,\cdots ,m+n,\end{array}
\right.
\end{eqnarray}

\noindent $\eta $ is an anisotropy parameter,
$a,b$ are indices running from 1 to $m+n$. 
For convenience, we denote 
 
\begin{equation}
\begin{array}{rclcrcl}
a(u)&=&sin(u+\eta )&,& b(u)&=&sin(u),\\
w(u)&=&sin(\eta -u)&,&c_{\pm}(u)&=&sin({\eta })e^{\pm iu}.
\end{array}
\end{equation}

\noindent This model is an $m+n$ state vertex model characterized by 
$m$ bosons and $n$ fermions. If $m=2$, $n=1$, this model reduces to 
the one studied by Foerster, Karowski [16] and Gonzalez-Ruiz [17]. The 
 $A_{m-1}$ vertex model studied by de Vega and Gonzalez-Ruiz [18] is just special case
 $n=0$. When $m=n=2$, we can get a new electronic strong interaction model
which is a generalization of the model proposed by Essler, Korepin and Schoutens  [22].

The R-matrix defined by (2,3) is a trigonometric solution of the
Yang-Baxter equation (1). The local transition matrix which is the
operator representation of Yang-Baxter equation 
is the $(m+n)\times (m+n)$
matrix $L(u)$ satisfying the following equation [19,20,21]:

\begin{eqnarray}
R_{12}(u-v)L_1(u)L_2(v)=L_2(v)L_1(u)R_{12}(u-v)
\end{eqnarray}

\noindent where $L_1(u)=L(u)\otimes 1$, $L_2(u)=1\otimes L(u)$. The 
standard row-to-row monodromy matrix for an $N\times N$ square lattice 
is defined by

\begin{eqnarray}
T(u)&=&L_N(u)\cdots L_1(u)\nonumber \\
&=&R_{0N}(u)\cdots R_{01}(u),
\end{eqnarray}

\noindent Throughout the paper, $L(u)$ is assumed to be in the  fundamental representation. 
$T(u)$ also fulfills the Yang-Baxter equation

\begin{eqnarray}
R_{12}(u-v)T_1(u)T_2(v)=T_2(v)T_1(u)R_{12}(u-v).
\end{eqnarray}

\noindent  The operator $T$ is an   $(m+n)\times (m+n)$ matrix of 
the operators acting in the quantun 
space $V_{n+m}^{\otimes N}$.

We can see that the R-matrix does not satisfy the Sklyanin's 
$P$- and $T$-symmetry, but fulfills $PT$ invariance

\begin{eqnarray}
P_{12}R_{12}(u)P_{12}=R_{12}^{t_1t_2}(u).
\end{eqnarray}

\noindent It also obeys the unitarity and cross-unitarity properties

\begin{eqnarray}
R_{12}(u)R_{21}(-u)=sin(u+\eta )sin(\eta -u)\cdot id,\\[5mm]
R_{12}^{t_1}(u)M_1R_{12}^{t_2}(-u-d\eta )M_1^{-1}=-sin(u)
sin(u+d\eta )\cdot id.
\end{eqnarray}

\noindent where $d=m-n$ and $M$ is a $(m+n)\times (m+n)$ matrix 

\begin{eqnarray}
M_{bc}&=&\delta _{bc}M_b, \nonumber \\[5mm]
M_b&=&\left\{ \begin{array}{ll} {\epsilon _d}e^{i2\eta (b-1)},&b\leq m\\[3mm]
                             {\epsilon _d}e^{i2\eta (2m-b)}. &b>m 
                             \end{array}
                             \right.
\end{eqnarray}

\noindent This can be verified by straightforward calculation. One can
also verify that $M$ is a symmetry matrix of the R-matrix:

\begin{eqnarray}
[M\otimes M,R_{12}(u)]=0
\end{eqnarray}

Now, we can use Mezincescue and  
Nepomechie's generalized formalism to construct integrable 
systems with open boundary conditions. In  our case, the reflection
equations take the following form [7]:

\begin{equation} 
R_{12}(u-v)K_1^-(u)R_{21}(u+v)K_2^-(v)=K_2^-(v)R_{12}(u+v)K_1^-(u)
R_{21}(u-v)
\end{equation}
\begin{eqnarray}
R_{12}(-u+v)K_1^+(u)^{t_1}M_1^{-1}R_{21}(-u-v-d\eta )
M_1K_2^+(v)^{t_2}\nonumber\\[2mm]
= K_2^+(v)^{t_2}M_1R_{12}(-u-v-d\eta )M_1^{-1}K_1^+(u)^{t_1}R_{21}(-u+v)
\end{eqnarray}

\noindent Obviously, there is an isomorphism between $K^+(u)$ and
$K^-(u)$. 

\begin{eqnarray}
\phi:K^-(u)\rightarrow K^+(u)=K^-(-u-{d\eta \over 2})^tM
\end{eqnarray}

Therefore, given a solution $K^-(u)$ of equation (13), we can also find a solution $K^+(u)$ of 
equation (14). But in a transfer
matrix of an integrable lattice, $K^-(u)$ and $K^+(u)$ need not 
satisfy equation (15). In this paper, we will take equation (15) to define $K^+$.  

After a long calculation, we find a solution of the reflection equation (13) 

\begin{equation}
K^-(u)=id
\end{equation}

\noindent Correspondingly, 

\begin{equation}
K^+(u)=M
\end{equation}

Taking  Sklyanin's formalism, the double-row monodromy matrix is 
defined as:

\begin{eqnarray}
U(u)=T(u)K^-(u)T^{-1}(-u)
\end{eqnarray}

\noindent where $T^{-1}(u)$ is the inverse of $T(u)$ in the auxiliary
and quantum spaces, which explicitly is:

\begin{eqnarray}
T^{-1}(u)&=&L_1^{-1}(u)\cdots L_N^{-1}(u)\nonumber \\
        &=&R_{01}^{-1}(u)\cdots R_{0N}^{-1}(u)
\end{eqnarray}

\noindent With the help of the Yang-Baxter equation (7) and the reflection equation (13),
one can prove that the double-row monodromy matrix satisfes the
reflection equation

\begin{eqnarray}
R_{12}(u-v)U_1(u)R_{21}(u+v)U_2(v)=U_2(v)R_{12}(u+v)U_1(u)R_{21}(u-v)
\end{eqnarray}

\noindent In this case, the transfer matrix is defined as:

\begin{eqnarray}
t(v)=trK^+(v)U(v).
\end{eqnarray}

\noindent Using the reflection equations (14,20) and the  
properties of the R-matrix (8-12), one can prove

\begin{equation}
[t(u),t(v)]=0.
\end{equation}

\noindent So the transfer matrix constitutes a one-parameter commutative
family which ensures the integrability of the model. As indicated by
Sklyanin, the transfer matrix is related to the hamiltonian of the quantum chain
with nearest neighbour interaction and boundary terms

\begin{eqnarray}
t'(0)=2trK^+(o)\frac1{sin(\eta) }H-2Ncot{\eta }
\end{eqnarray}

\noindent From equation (21), one can derive the explicit expression of the
 hamiltonian, which is omitted here because it is not used in the following discussion.
In order to compare it with the $SU_q(2|1)$ supersymmetric t-J model, we give the
 hamiltonian under $m=1$, which is defined:

\begin{eqnarray}
H&=&P\{ {\sum_{j=1}^{N-1}}{\sum_s}(c^{\dagger }_{j,s}c_{j+1,s}
+c^{\dagger }_{j+1,s}c_{j,s})\} P\nonumber \\[2mm]
& &+\sum_{j=1}^{N-1}cos\eta\left(\sum_{a=1}^m n_{j,a}n_{j+1,a}+cos\eta\cdot(n_j+n_{j+1})-
cos\eta\cdot(n_jn_{j+1})\right)  \nonumber \\[2mm]
& &+\sum_{j=1}^{N-1}\left(\sum_{\alpha\in \Delta_+}(S_j^{\alpha}S_{j+1}^{-\alpha}
+S_j^{-\alpha}S_{j+1}^{\alpha})+isin\eta\cdot(n_j-n_{j+1})\right)\nonumber \\[2mm]
& &+\sum_{j=1}^{N-1}isin\eta\cdot(\sum_{a<b}n_{j,a}n_{j+1,b}-\sum_{a>b}n_{j,a}n_{j+1,b}),
\end{eqnarray}

\noindent where $c_{js}^{\dagger }$ ($c_{js}$)  creates (annihilates) an
electron with spin component $s$,$s=1,2,\cdots ,n+m-1$ located j-th site. $n_j$ is the 
density operator,$S_j$ is the spin matrix at site $j$. $\Delta_+$ denotes the set of positive
roots of the $su(m)$ algebra. $P$ is a operator  projecting out
doubly occupied states. The constraint 
is that  more than one electron on  
each site is strictly prohibited. As we know [16,17], this
hamiltonian is not hermitean, but it possesses real eigenvalues. We
will show the hamiltonian to be $SU_q(n|m)$ invariant.

\section{Nested Bethe ansatz for open boundary conditions}

The graded vertex model with periodic boundary condition was investigated 
by de Vega and Lopes [20,21].  Based upon the Yang-Baxter equation, they obtain
the Bethe Ansatz equations by using the nested Bethe ansatz method (periodic case).
In this section, we want to generalize the nested Bethe ansatz method  
to solve the eigenvalue problem of the transfer matrix (21). In this case, 
the operator commutative relations are ruled by the reflection equation instead of 
the Yang-Baxter equation. 

As we know, the double-row monodromy matrix satisfies the reflection
equation. It is convenient to denote $u_-=u-v$, $u_+=u+v$. We rewrite
the equation (20) in the component form:

\begin{eqnarray}
R_{12}(u_-)^{a_1a_2}_{c_1c_2}U(u)_{c_1d_1}R_{21}(u_+)^{d_1c_2}_{b_1d_2}
U(v)_{d_2b_2}\nonumber \\[4mm]
=U(v)_{a_2c_2}R_{12}(u_+)^{a_1c_2}_{c_1d_2}U(u)_{c_1d_1}
R_{21}(u_-)^{d_1d_2}_{b_1b_2}
\end{eqnarray}

\noindent where the repeated indices  sum over 1 to $m+n$.
Next, we introduce a set of notations for convenience:

\begin{eqnarray}
A(v)&=&U(v)_{11},\nonumber \\
B_a(v)&=&U(v)_{1a},\nonumber \\
C_a(v)&=&U(v)_{a1},\nonumber \\
D_{ab}(v)&=&U(v)_{ab}, 2\leq a,b\leq m+n.
\end{eqnarray}

\noindent From equation (25) we will find the commutation relations. In order
to simplify these relations,  we introduce new operators:

\begin{eqnarray}
\tilde{D}_{ab}(v)=D_{ab}(v)-{\delta _{ab}}\frac {R_{12}(2v)^{a1}_{1a}}
{R_{12}(2v)^{11}_{11}}A(v)
\end{eqnarray}

\noindent Considering the vertex model defined by equations (2,3), we rewrite
equation (27) in an explicit form:

\begin{eqnarray}
\tilde{D}_{ab}(v)=\left\{ \begin{array}{ll} 
D_{ab}(v)-\delta _{ab}\displaystyle\frac {c_+(2v)}{a(2v)}A(v),&m\not=0.\\[3mm]
D_{ab}(v)-\delta _{ab}\displaystyle\frac {c_+(2v)}{w(2v)}A(v),&m=0.
\end{array}\right.
\end{eqnarray}

\noindent After some tedious calculation, we have found the commutation
relations between $A(v)$, $\tilde{D}_{ab}(v)$ and $B_a(u)$ $(a,b=2,
\cdots,m+n)$. The final results take the form (see Appendix A)

\begin{eqnarray}
A(v)B_b(u)&=&\frac {a(u-v)b(u+v)}{a(u+v)b(u-v)}B_b(u)A(v)\nonumber \\[3mm]
& &-\frac {c_+(u-v)b(2u)}{a(2u)b(u-v)}B_b(v)A(u)\nonumber \\[3mm]
& &-\frac {c_-(u+v)}{a(u+v)}B_c(v)\tilde{D}_{cb}(u),
\end{eqnarray}
 
\begin{eqnarray}
\tilde{D}_{a_1b_1}(u)B_{b_2}(v)
&=&\,\,{\frac{R_{12}(u+v+\eta )^{a_1c_2}_{c_1d_2}}{b(u-v)b(u+v+\eta)}}
    R_{21}(u-v)^{d_1d_2}_{b_1b_2} B_{c_2}(v)\tilde{D}_{c_1d_1}(u)\nonumber \\[3mm]
& &-{\frac {c_+(u-v)}{b(2u+\eta )b(u-v)}}R_{12}(2u+\eta )^{a_1d_1}_{d_2b_1}
   B_{d_1}(u)\tilde{D}_{d_2b_2}(v)\nonumber \\[3mm]
& &+{1\over {b(2u+\eta )}}{\frac {c_+(u+v)b(2v)}{a(u+v)a(2v)}}
R_{12}(2u+\eta )^{a_1d_2}_{b_2b_1}B_{d_2}(u)A(v).
\end{eqnarray}

\noindent All indices take values fron 2 to $m+n$, and the 
repeated indices  sum over 2 to $m+n$. The commutation 
relations presented above are only applicable to the cases $m\geq 1$.
If m=0, the commutation relations change to the following form:

\begin{eqnarray}
A(v)B_b(u)&=&{\frac {w(u-v)b(u+v)}{w(u+v)b(u-v)}}B_b(u)A(v)\nonumber \\[3mm]
& &-\frac {c_+(u-v)b(2u)}{w(2u)b(u-v)}B_b(v)A(u)\nonumber \\[3mm]
& &-\frac {c_-(u+v)}{w(u+v)}B_c(v)\tilde{D}_{cb}(u)
\end{eqnarray}

\begin{eqnarray}
\tilde{D}_{a_1b_1}(u)B_{b_2}(v)
&=&\,\,{\frac{R_{12}(u+v-\eta )^{a_1c_2}_{c_1d_2}}{b(u-v)b(u+v-\eta)}}
       R_{21}(u-v)^{d_1d_2}_{b_1b_2}B_{c_2}(v) \tilde{D}_{c_1d_1}(u)\nonumber \\[3mm]
& &-{\frac {c_+(u-v)}{b(2u-\eta )b(u-v)}}R_{12}(2u-\eta )^{a_1d_1}_{d_2b_1}
B_{d_1}(u)\tilde{D}_{d_2b_2}(v)\nonumber \\[3mm]
& &+{1\over b(2u-\eta )}{\frac {c_+(u+v)b(2v)}{w(u+v)w(2v)}}
R_{12}(2u-\eta )^{a_1d_2}_{b_2b_1}B_{d_2}(u)A(v)
\end{eqnarray}

\noindent The rule for indices is the same as the one in equations (29,30).

It is easy to find the so-called local vacuum $e_i^+$. We call the 
direct product of local vacuum a reference state or vacuum state. It
takes the form:

\begin{eqnarray}
|vac>=\prod^{\otimes N} (1,0,\cdots ,0)^t,
\end{eqnarray}

\noindent where $t$ denotes the transposition. One can find

\begin{eqnarray}
A(u)|vac>&=&\alpha (u)|vac>,\nonumber \\
C_a(u)|vac>&=&0,\nonumber \\
B_a(u)|vac>&\ne &0,\nonumber \\
\alpha (u)&=&[R(u)^{11}_{11}]^N[R^{-1}(-u)^{11}_{11}]^N.
\end{eqnarray}

\noindent Next, let us calculate the action of $\tilde{D}_{ab}(u)$ on the 
vacuum state. We first recall the definition of $D_{ab}(u)$, and find

\begin{eqnarray}
D_{ab}(u)|vac>=T(u)_{a1}T^{-1}(-u)_{1b}|vac>+T(u)_{ac}T^{-1}(-u)_{cb}|vac>.
\end{eqnarray}

\noindent The contribution  of the first term  can 
not be calculated directly. We will use the following method to find it.
Taking $v=-u$ in the Yang-Baxter equation, we can get:

\begin{eqnarray}
T_2^{-1}(-u)R_{12}(2u)T_1(u)=T_1(u)R_{12}(2u)T^{-1}_2(-u)
\end{eqnarray}

\noindent Taking special indices in this relation and 
applying both sides of this relation to the vacuum state, we find:

\begin{eqnarray}
T(u)_{a1}T^{-1}(-u)_{1b}|vac>={\frac {c_+(2u)}{R(2u)^{11}_{11}}}
\left( \delta _{ab}\alpha (u)-T(u)_{ac}T^{-1}(-u)_{cb}\right) |vac>.
\end{eqnarray}

\noindent Substituting this relation to eq.(35), we have the result:

\begin{eqnarray}
D_{ab}(u)|vac>=\delta _{ab}\{ {\frac {c_+(2u)}{R(2u)_{11}^{11}}}\alpha (u)
+\left( 1-{\frac {c_+(2u)}{R(2u)^{11}_{11}}}\right) 
b^N(u)\tilde{b}^N(-u)\} |vac>.
\end{eqnarray}                                     

\noindent So we have

\begin{eqnarray}
\tilde{D}_{ab}(u)|vac>=\delta _{ab}\left( 1-{\frac {c_+(2u)}
{R(2u)^{11}_{11}}}\right)
b^N(u)\tilde{b}^N(-u)|vac>,
\end{eqnarray}

\noindent where $\tilde{b}(u)=R^{-1}(u)^{ab}_{ab}, a\ne b.$
In conclusion, the results of the action of $A,B_a,C_a$ and 
$\tilde{D}_{ab}$ on the vacuum state are listed as:

\begin{eqnarray}
A(u)|vac>&=&[R(u)^{11}_{11}]^N[R^{-1}(-u)^{11}_{11}]^N|vac>
=\alpha (u)|vac>\nonumber \\
\tilde{D}_{ab}(u)|vac>
&=&\delta _{ab}\left( 1-{\frac {c_+(2u)}{R(2u)^{11}_{11}}}
\right) b^N(u)\tilde{b}^N(-u)|vac>=\delta _{ab}\beta (u)|vac>\nonumber \\
C_a(u)|vac>&=&0\nonumber \\
B_a(u)|vac>&\ne &0.
\end{eqnarray}
 
\noindent Note that the action of $B_a(u)$ on the vacuum state is not 
proportional to the vacuum state.

We show that the eigenvectors of transfer matrix $t(u)$ can be constructed
by repeatedly applying operators $B_{b_i}(v_i)$ on the vacuum state  

\begin{eqnarray}
\Psi (v_1,\cdots ,v_L)=B_{b_1}(v_1)\cdots B_{b_L}(v_L)|vac>F^{b_1\cdots b_L}
\end{eqnarray}
\noindent Before using the Bethe ansatz method, let us introduce a set
of notatons that will be used in the following.  We denote 
 
\begin{equation}
R_{21}(u)^{ab}_{cd}/R(u)^{11}_{11}=\tilde{R}_{21}(u)^{ab}_{cd}.
\end{equation}

\noindent So from Appendix A, the commutation relations between B's take
the form

\begin{eqnarray}
B_{b_1}(u_1)B_{b_2}(u_2)=\tilde{R}_{12}(u_1-u_2)^{d_2d_1}_{b_2b_1}
B_{d_2}(u_2)B_{d_1}(u_1).
\end{eqnarray}

\noindent By repeatedly using this relation, we can commute $B(v_k)$ 
with $B(v_{k-1}), \cdots ,B(v_1)$, respectively.

\begin{eqnarray}
& &B_{b_1}(v_1)\cdots B_{b_L}(v_L)|vac>\nonumber \\
&=&S(v_k, \{ v_i\} )^{d_1\cdots d_L}_{b_1\cdots b_L}B_{d_1}(v_k)
B_{d_2}(v_1)\cdots B_{d_k}(v_{k-1})B_{d_{k+1}}(v_{k+1})\cdots 
B_{d_L}(v_L)|vac>
\end{eqnarray}

\noindent where 

\begin{eqnarray}
& &S(v_k, \{ v_i\} )^{d_1\cdot d_L}_{b_1\cdots b_L}\nonumber \\
&=&\prod_{j=1+k}^{L}
\delta _{b_jd_j}\tilde{R}_{12}(v_1-v_k)^{d_1d_2}_{c_2b_1}
\tilde{R}_{12}(v_2-v_k)^{c_2d_3}_{c_3b_2}\cdots 
\tilde{R}_{12}(v_{k-1}-v_k)^{c_{k-1}d_k}_{b_kb_{k-1}}
\end{eqnarray}

\noindent Here $d_1$ and $b_k$ are considered as the  "auxiliary space" indices, 
$b_1,\cdots,b_{k-1},b_{k+1},\cdots ,b_L$ and  $d_2,\cdots ,d_L$ are the  "quantum space" indices.
Notice that in the six vertex model, the B's are commutable with
each other. So one can use the symmetric argument that $v_k$ and $v_1$
are equivalent to each other. Now, we know that the B's are not 
commutable with each other, but the relation (44) ensures that 
we can also use something like the symmetric argument.  
 Actually, we can see from this relation that $v_k$ and $v_1$
are in an equivalent position if we omit the function $S$.

In the following, we  deal with the case of $R(u)^{11}_{11}=a(u)$.
It is convenient to introduce the notation:

\begin{eqnarray}
L^{(1)}(\tilde{v}_1,\tilde{v}_i)^{a_1b_1}_{a_2b_2}=
R_{12}(\tilde{v}_1+\tilde{v}_i)^{a_1b_1}_{a_2b_2},\nonumber
\end{eqnarray}

\begin{eqnarray}
[L^{(1)}(\tilde{v}_1,\tilde{v}_i)^{-1}]^{a_1b_1}_{a_2b_2}=
{\frac {R_{21}(-\tilde{v}_1-\tilde{v}_i)^{a_1b_1}_{a_2b_2}}
{a(v_1+v_i)a(-v_1-v_i)}},\nonumber 
\end{eqnarray}

\begin{eqnarray}
& &R_{12}(\tilde{v}_1+\tilde{v}_L)^{c_{L-1}d_L}_{q_Lp_L}
R_{21}(\tilde{v}_1-\tilde{v}_L)^{q_Lp_L}_{q_{L-1}b_L}
\beta _{q_L}(v_1)\nonumber \\
&=&{\frac 1{a(\tilde{v}_1-\tilde{v}_L)a(\tilde{v}_L-\tilde{v}_1)}}
[L^{(1)}(\tilde{v}_1,\tilde{v}_L)\beta (\tilde{v}_1)
L^{(1)}(-\tilde{v}_1,\tilde{v}_L)^{-1}]^{c_{L-1}d_L}_{q_{L-1}b_L}.
\end{eqnarray}

\noindent Here we have used the unitarity properties of R matrix, and
$\tilde{v}_i=v_i+\eta /2$, $\left( \beta(\tilde{v}_1)\right) _{ab}
=\delta _{ab}\beta(\tilde{v}_1)$. 

Now, let us evaluate the action of $A(u)$ on  $\Psi $. 
Following the algebraic Bethe ansatz method, many terms will appear when we 
move  $A(u)$ from the left hand side to the right hand side of $B_a$'s.
They can be classified in two
types: wanted and unwanted terms. The wanted terms in $A(u)\Psi $
can be obtained by repeatedly using the first term in relation (29), 
the unwanted terms arise from the second and third terms in relation (29), 
they are the types that $v_k$ is replaced by $u$. One unwanted term  
where $B(v_1)$ is replaced by $B(u)$ can be obtained  by using first the second and 
third terms in relation (29), then repeatedly using the first terms
in relation (29) and (30). Using this results we can obtain the 
general unwanted term where $B(v_k)$ is replaced by $B(u)$.
 So we can find  the action of $A(u)$ on  $\Psi $

\begin{eqnarray}
& & A(u)B_{b_1}(v_1)\cdots B_{b_L}(v_L)|vac>F^{b_1\cdots b_L}\nonumber \\[3mm]
&=& \,\,{\prod _{j=1}^{L}}{\frac {a(v_j-u)b(v_j+u)}{b(v_j-u)a(v_j+u)}}
    \alpha (u)\cdot B_{b_1}(v_1)\cdots B_{b_L}(v_L)|vac>
    F^{b_1\cdots b_L}\nonumber \\[3mm]
& & +\,{\sum_{k=1}^L}{\frac {-c_+(v_k-u)b(2v_k)}{a(2v_k)b(v_k-u)}}
     {\prod _{j=1,\ne k}^L}{\frac {a(v_j-v_k)b(v_j+v_k)}{a(v_j+v_k)b(v_j-v_k)}}
     \alpha (v_k)F^{b_1\cdots b_L}\nonumber \\[3mm]
& &  \cdot B_{d_1}(u)B_{d_2}(v_1)\cdots B_{d_k}(v_{k-1})B_{d_{k+1}}(v_{k+1})
     \cdots B_{d_L}(v_L)|vac>
     S(v_k,\{ v_i\} )^{d_1 \cdots d_L}_{b_1 \cdots b_L}\nonumber \\[3mm]
& &+\,{\sum _{k=1}^L}{\frac {c_-(v_k+u)}{a(v_k+u)}}
    {\prod  _{j=1,\ne k}^L}{\frac {a(v_j-v_k)a(v_k-v_j)}{b(v_j+v_k+\eta )
    b(v_k-v_j)}}\beta(v_k)S(v_k,\{ v_i\} )^{c_1\cdots c_L}_{b_1\cdots b_L}   \\[3mm]
& & \,\,[L^{(1)}(\tilde{v}_k,\tilde{v}_1) \cdots L^{(1)}(\tilde{v}_k,\tilde{v}_{k-1})
    L^{(1)}(\tilde{v}_k,\tilde{v}_{k+1})\cdots 
    L^{(1)}(\tilde{v}_k,\tilde{v}_L)\nonumber \\[3mm]
& & \,\,\cdot L^{(1)}(-\tilde{v}_k,\tilde{v}_L)^{-1}\cdots 
    L^{(1)}(-\tilde{v}_k,\tilde{v}_{k-1})^{-1}L^{(1)}(-\tilde{v}_k,
    \tilde{v}_{k+1})^{-1}\cdots L^{(1)}(-\tilde{v}_k,\tilde{v}_L)^{-1}]
    ^{d_1\cdots d_L}_{c_1\cdots c_L}\nonumber \\[3mm]
& & \,\,\cdot B_{d_1}(u)B_{d_2}(v_1)\cdots B_{d_k}(v_{k-1})B_{d_{k+1}}(v_{k+1})
     \cdots B_{d_L}(v_L)|vac> F^{b_1\cdots b_L}\nonumber
\end{eqnarray}

Recalling the definition of the transfer matrix,  we rewrite the transfer matrix as:

\begin{eqnarray}
t(u)&=&{\sum _{a=1}^{m+n}}K^+_a(u)U(u)_{aa}\nonumber \\
    &=&\sum_{a=2}^{m+n}K^+_a(u)\tilde{D}_{aa}(u)+
       +\left\{\sum_{a=2}^{m+n}K^+_a(u)\frac{c_+(2u)}{a(2u)}+K^+_1(u)\right\}A(u)\nonumber \\
    &=&\sum_{a=2}^{m+n}K^+_a(u)\tilde{D}_{aa}(u)+\frac{sin(2u+d\eta)}
        {sin(2u+\eta)}e^{i(d-1)\eta}A(u)
\end{eqnarray}

\noindent Before calculating the action of the transfer matrix on $\Psi$,  we should evaluate 
the action of $K^+_a(u)\tilde{D}_{aa}(u)$ on it, which reads
\begin{eqnarray}
& &\sum_{a=2}^{m+n}K^+_a(u)\tilde{D}_{aa}(u)
B_{b_1}(v_1)\cdots B_{b_L}(v_L)|vac>F^{b_1\cdots b_L}\nonumber \\
&=&\sum _{a=2}^{m+n}K^+_a(u)F^{b_1\cdots b_L}\beta(u)
\prod _{j=1}^L
{\frac 1{b(u-v_j)b(u+v_j+\eta )}}\nonumber \\
& &\cdot R_{12}(u+v_1+\eta )^{ad_1}_{p_1q_1}R_{21}(u-v_1)^{s_1q_1}_{ab_1}
R_{12}(u+v_2+\eta )^{p_1d_2}_{p_2q_2}R_{21}
(u-v_2)^{s_2q_2}_{s_1b_2}\nonumber \\
& &\cdots R_{12}(u+v_L+\eta )^{p_{L-1}d_L}_{p_Lq_L}R_{21}(u-
v_L)^{s_Lq_L}_{s_{L-1}b_L}\nonumber \\
& &B_{d_1}(v_1)B_{d_2}(v_2)\cdots B_{d_L}(v_L)\delta_{p_ns_n}|vac>+ u.t.
\end{eqnarray}
where $u.t.$ stands for unwanted term.
Using the definition of $L^{(1)}(\tilde{u},\tilde{v}_i)$ and its inverse, we rewrite
relation  (49) as:

\begin{eqnarray}
\cdots&=&\sum_{a=2}^{m+n}K^+_a(u)F^{b_1\cdots b_L}\prod_{j=1}^L
         \frac{a(u-v_j)a(v_j-u)}{b(u-v_j)b(u+v_j+\eta)}\beta(u)\nonumber \\
      & &\left\{\left(T^{(1)}(\tilde{u},\{\tilde{v}_i\})T^{(1)}(-\tilde{u},\{\tilde{v}_i\})^{-1}
         \right)^{d_1\cdots d_L}_{b_1\cdots b_L}\right\}_{aa}\nonumber \\
      & &\cdot B_{d_1}(v_1)B_{d_2}(v_2)\cdots B_{d_L}(v_L)|vac>  
         + u.t.
\end{eqnarray}

\noindent Here

\begin{eqnarray}
& &\left\{ \left( T^{(1)}(\tilde{u},\{ \tilde{v}_i\} )
T^{(1)}(-\tilde{u},\{ \tilde{v}_i\} )^{-1}
\right) ^{d_1\cdots d_L}_{b_1\cdots b_L}\right\} _{aa}\nonumber \\
&=&\left\{ \left( L^{(1)}(\tilde{u},\tilde{v}_1)\cdots 
L^{(1)}(\tilde{u},\tilde{v}_L)L^{(1)}(-\tilde{u},\tilde{v}_L)^{-1}
\cdots L^{(1)}(-\tilde{u},\tilde{v}_1)^{-1}\right)^{d_1\cdots d_L}_{b_1\cdots b_L}\right\}_{aa}
\end{eqnarray}

\noindent As mentioned above $a$ is the index of the auxiliary space and 
$b_i,d_i$ are the indices of the quantum space.
The unwanted terms in equation (49) take two forms. After some long tedious calculation
based upon the similar considerations as in the $A(u)$ case,
we can get the following expression After long tedious calculation

\begin{equation}
\begin{array}{rcl}
u.t. &=& \sum^{L}_{k=1}S(u_k,\{v_i\})^{d_1\cdots d_L}_{b_1\cdots b_L}F^{b_1\cdots b_L}
           \cdot \alpha(v_k) e^{id\eta}  \\[5mm]
     & &\,\displaystyle\frac{c_+(u+v_k)b(2v_k)b(2u+d\eta)}{a(u+v_k)a(2v_k)b(2u+\eta)}
          \prod_{j=i,\neq k}^L \frac{a(v_j-v_k)b(v_j+v_k)}{a(v_j+v_k)b(v_j-v_k)}\\[5mm]
     & &\,\cdot B_{d_1}(u)B_{d_2}(v_1)\cdots B_{d_k}(v_{k-1})B_{d_{k+1}}(v_{k+1})
         \cdots B_{d_L}(v_L)|vac>\\[5mm]
     & & + \sum^{L}_{k=1}S(u_k,\{v_i\})^{c_1\cdots c_L}_{b_1\cdots b_L}F^{b_1\cdots b_L}
            \cdot \beta(v_k)  e^{id\eta}\\[5mm]
     & &\,\displaystyle\frac{-c_+(u-v_k)b(2u+d\eta)}{b(u-v_k)b(2u+\eta)} 
        \prod_{j=i,\neq k}^L \frac{a(v_j-v_k)a(v_k-v_j)}{b(v_k-v_j)b(v_j+v_k+\eta)}\\[5mm]
     & & \,\left(L^{(1)} (\tilde{v}_k,\tilde(v)_1)
          \cdots L^{(1)}(\tilde{v}_k,\tilde{v}_{k-1}) 
      L^{(1)}(\tilde{v}_k,\tilde{v}_{k+1})\cdots\right. \\[5mm]
     & & \cdots L^{(1)}(\tilde{v}_k,\tilde{v}_L) L^{(1)}(-\tilde{v}_k,\tilde{v}_L)^{-1}
          \cdots L^{(1)}(-\tilde{v}_k,\tilde{v}_{k+1})^{-1}\\[5mm]
    & &\left. L^{(1)}(-\tilde{v}_k, \tilde{v}_{k-1})^{-1}\cdots 
             L^{(1)}(-\tilde{v}_k,\tilde{v}_1)^{-1}\right)
             ^{d_1\cdots d_L}_{c_1\cdots c_L}\\[5mm]
    & & \,B_{d_1}(u)B_{d_2}(v_1)\cdots B_{d_k}(v_{k-1})B_{d_{k+1}}(v_{k+1})
                        \cdots B_{d_L}(v_L)|vac>.
\end{array}
\end{equation}

In order to simplify equations (47) and (52), we need an 
important relation
\begin{eqnarray}
& & \left\{L^{(1)} (\tilde{v}_k,\tilde(v)_1)\cdots L^{(1)}(\tilde{v}_k,\tilde{v}_{k-1})
    L^{(1)}(\tilde{v}_k,\tilde{v}_{k+1})\cdots \cdots L^{(1)}(\tilde{v}_k,\tilde{v}_L)\right.
    \nonumber \\
& & \cdot L^{(1)}(-\tilde{v}_k,\tilde{v}_L)^{-1} \cdots L^{(1)}(-\tilde{v}_k,\tilde{v}_{k+1})^{-1}
     L^{(1)}(-\tilde{v}_k, \tilde{v}_{k-1})^{-1}\cdots \nonumber \\
& &\left.\cdot L^{(1)}(-\tilde{v}_k,\tilde{v}_1)^{-1}\right\}^{d_1\cdots d_L}_{c_1\cdots c_L}
   S(v_k,\{v_i\})^{c_1\cdots c_L}_{b_1\cdots b_L}\nonumber  \\
& =&\, \frac{sin(\eta)}{sin(2v_k+d\eta)}e^{-id\eta}S(v_k,\{v_i\})^{c_1\cdots c_L}_{b_1\cdots b_L}
       \tau^{(2)}(\tilde{v_k},\{\tilde{v_i}\})^{c_1\cdots d_L}_{b_1\cdots b_L}
\end{eqnarray}
where
\begin{eqnarray}
& &\,\, \tau^{(2)}(u,\{\tilde{v}_i\})^{c_1\cdots d_L}_{b_1\cdots b_L} \nonumber \\
&=&\sum_{a=2}^{m+n}K^+_a(u)\left\{\left( L^{(1)} (u,\tilde{v}_1)\cdots
 L^{(1)}(u,\tilde{v}_L)\right.\right.  \nonumber \\
& & \left.\left.\cdot L^{(1)}(-u,\tilde{v}_L)^{-1} \cdots L^{(1)}(-u,\tilde{v}_{1})^{-1}
    \right)_{b_1\cdots b_L}^{c_1\cdots c_L}\right\}_{aa} \nonumber \\
&=&\sum_{a=2}^{m+n}K^+_a(u)\left\{\left( T^{(1)} (u,\{\tilde{v}_i\})T^{(1)}(-u,\{\tilde{v}_i\})^{-1}\right)
_{b_1\cdots b_L}^{c_1\cdots c_L}\right\}_{aa}
\end{eqnarray}
 
Using the equations (48) and (53), we then obtain the action of $t(u)$ on  $\Psi$

\begin{equation}
\begin{array}{rcl}
t(u)\Psi&=&\displaystyle\alpha(u)e^{i(d-1)\eta}\frac{sin(2u+d\eta)}{sin(2u+\eta)}
           \prod_{j=1}^L\frac{a(v_j-u)b(v_j+u)}{b(v_j-u)a(v_j+u)}\\[5mm]
        & & F^{b_1\cdots b_L}B_{b_1}(v_1)\cdots B_{b_L}(v_L)|vac> \\[5mm]
        & &\displaystyle+ \prod_{j=1}^L\frac{a(v_j-u)a(u-v_j)}{b(u-v_j)b(v_j+u+\eta)}\beta(u)
           \tau^{(2)}(\tilde{u},\{\tilde{v}_i\})^{d_1\cdots d_L}_{b_1\cdots b_L}\\[5mm]
         & & F^{b_1\cdots b_L} B_{d_1}(v_1)\cdots B_{d_L}(v_L)|vac> \\[5mm]
        & &\displaystyle+\sum_{k=1}^L\left(\frac{-c_+(v_k-u)}{b(v_k-u)}e^{-i\eta}
           +\frac{c_+(v_k+u)}{a(v_k+u)}\right)\frac{sin(2u+d\eta)b(2v_k)}{sin(2u+\eta)a(2v_k)}e^{id\eta}
           \\[5mm]
        & &\displaystyle\cdot\prod_{j=1,\neq k}^{L}
           \frac{a(v_j-v_k)b(v_j+v_k)}{b(v_j-v_k)a(v_j+v_k)}\alpha(v_k)S(v_k,\{v_i\})^{d_1\cdots
           d_L}_{b_1\cdots b_L}F^{b_1\cdots b_L}\\[5mm]
        & &\displaystyle\,\cdot B_{d_1}(u)b_{d_2}(v_1)\cdots B_{d_k}(v_{k-1})B_{d_{k+1}}(v_{k+1})
           \cdots B_{d_L}(v_L)|vac>\\[5mm]
        & &\displaystyle-\sum_{k=1}^L\left(\frac{c_-(u+v_k)}{a(u+v_k)}e^{-i\eta}+\frac{c_+(u-v_k)}
           {b(u-v_k)}\right)\frac{sin(2u+d\eta)}{sin(2u+\eta)}\\[5mm]
         & &\displaystyle\frac{sin(\eta)}{sin(2v_k+\eta)}\prod_{j=1,\neq k}^L\frac{a(v_k-v_j)a(v_j-v_k)}
             {b(v_k-v_j)b(v_j+v_k+\eta)}\\[5mm]
         & & S(v_k,\{v_i\})^{d_1\cdots d_L}_{c_1\cdots c_L}
           \tau^{(2)}(\tilde{v}_k,\{\tilde{v}_i\})^{c_1\cdots c_L}_{b_1\cdots b_L}
          F^{b_1\cdots b_L} \\[2mm]
          & &B_{d_1}(u)b_{d_2}(v_1)\cdots 
               B_{d_k}(v_{k-1})B_{d_{k+1}}(v_{k+1}) \cdots B_{d_L}(v_L)|vac>
\end{array}
\end{equation}

From the above equation, one can see that the function $\Psi$ is not the eigenstate of $t(u)$ unless
$F$'s are the eigenstates of $\tau^{(2)}$ and the sum of the third and the fourth term in the above
equation is zero, which will give a restriction on the $L$ spectrum parameters $\{v_i\}$. 
 So, we have the following results:

If $F$ is the eigenstate of $\tau^{(2)}$ with  the eigenvalue $\Lambda^{(2)}$ satisfing
equation (57), then $\Psi$ is the eigenstate of $t(u)$ with  the eigenvalue $\Lambda^{(1)}$,

\begin{equation}
\begin{array}{rcl}
\Lambda^{(1)}(u)&=&\displaystyle\alpha(u)e^{i(d-1)\eta}\frac{sin(2u+d\eta)}{sin(2u+\eta)}
                   \prod_{j=1}^L\frac{a(v_j-u)b(v_j+u)}{b(v_i-u)a(v_j+u)} \\[2mm]
                & &\displaystyle\,+ \beta(u) \prod_{j=1}^L\frac{a(v_j-u)a(u-v_j)}{b(u-v_j)b(v_j+u+\eta)}
                   \Lambda^{(2)}(u,\{v_i\})
\end{array}
\end{equation}
\noindent where
\begin{equation}
\begin{array}{rcl}
\tau^{(1)}(u,\{v_i\})F&=&\Lambda^{(2)}(u,\{v_i\})F\\[2mm]
\Lambda^{(2)}(v_j,\{v_i\})&=&\displaystyle\frac{\alpha(v_k)b(2v_k)sin(2v_k+d\eta)}
                             {\beta(v_k)a(2v_k)sin(\eta)} 
                             e^{i(d-1)\eta}\prod_{j=1,\neq k}^L\frac{b(v_j+v_k)}{a(v_k-v_j)}
\end{array}
\end{equation}

Therefore, the diagonalization of $t(u)$ is reduced to finding the eigenvalue of $\tau^{(2)}$. The
explicit expression of $\tau^{(2)}$ (see equation (55)) implies that $\tau^{(2)}$ can be considered as
the transfer matrix of an $L$-sites quantum chain, in which every spin takes $m+n-1$ values. The
related Yang-Baxter equation is the same as the one of $t(u)$, exception $R$ being an 
$(m+n-1)^2\times (m+n-1)^2$ matrix. Hence, we can use 
the same method to find the eigenvalue of $\tau^{(2)}$. Repeating the procedure
$m$ times, one can reduce to a subsystem $\tau^{(m+1)}$ which is an $n\times n$ matrix in auxiliary 
space. The related Yang-Baxter equation is also defined by equation (2), but one should notice that
in this case all $\epsilon_a=-1$ due to $m=0$.  In order to diagonalize $\tau^{(m+1)}$, we need 
the definition of $\tilde{D}$ by the second equation (28). The elements of $T^{(m)}$ satisfy
equations (31) and (32). Following the same procedure, one can further reduce the $\tau^{(m+1)}$
 into the $\tau^{(m+2)}$ subsystem.
The late has the same structure as the former. In this case 
 one finaly obtains the eigenvalue of $\tau^{(m+n-1)}$.
This is the well-known nested Bethe Ansatz. Because the wave-functions are not 
needed in this paper, we omit them  
here. The eigenvalue and the constraint on the spectral parameters read as

\begin{equation}
\begin{array}{rcl}
\Lambda^{(k)}(u,\{v_i^{(k-1)}\},\{v_i^{(k)}\})&=&\displaystyle\alpha^{(k)}(u,\{v_i^{(k-1)}\})e^{i(d+k-2)\eta}
                   \frac{sin(2u+d\eta)}{sin(2u+k\eta)}\\[5mm]
                & & \displaystyle\cdot \prod_{j=1}^{P_k}\frac{a(v_j^{(k)}-u)b(u+v_j^{(k)}+(k-1)\eta)}
                   {b(v_j^{(k)}-u)a(u+v_j^{(k)}+(k-1)\eta)} \\[5mm]
               & &+\displaystyle \beta^{(k)}(u,\{v_i^{(k-1)}\}) \prod_{j=1}^{P_k}
                  \frac{a(v_j^{(k)}-u)a(u-v_j^{(k)})}{b(u-v_j^{(k)})b(u+v_j^{(k)}+k\eta)} \\[2mm]
                & & \cdot \Lambda^{(k+1)}(u,\{v_i^{(k)}\},\{v_i^{(k+1)}\}) \\[5mm]
                & &      ( 1\leq k \leq m)
\end{array}
\end{equation}

\begin{equation}
\begin{array}{rcl}
\Lambda^{(k)}(u,\{v_i^{(k-1)}\},\{v_i^{(k)}\})
       &=&\displaystyle\alpha^{(k)}(u,\{v_i^{(k-1)}\})e^{i(d+2m-k)\eta}
                   \frac{sin(2u+d\eta)}{sin(-2u+(k-2m)\eta)}\\[5mm]
                & &\cdot \displaystyle \prod_{j=1}^{P_k}\frac{w(v_j^{(k)}-u)b(u+v_j^{(k)}+(2m-k+1)\eta)}
                   {b(v_j^{(k)}-u)w(u+v_j^{(k)}+(2m-k+1)\eta)} \\[5mm]
               & &\displaystyle\,+ \beta^{(k)}(u,\{v_i^{(k-1)}\}) \prod_{j=1}^{P_k}
                  \frac{a(v_j^{(k)}-u)a(u-v_j^{(k)})}{b(u-v_j^{(k)})b(u+v_j^{(k)}+(2m-k)\eta)} \\[5mm]
               & &\cdot \Lambda^{(k+1)}(u,\{v_i^{(k)}\},\{v_i^{(k-1)}\}) \\[5mm]
                & &      ( m+1\leq k \leq n)
\end{array}
\end{equation}
\noindent and
\begin{equation}
\begin{array}{rcl}
\Lambda^{(k+1)}(v_l^{(k)},\{v_i^{(k)}\},\{v_i^{(k+1)}\})
  &=&\displaystyle\frac{\alpha^{(k)}(v^{(k)}_l,\{v_i^{(k-1)}\})}{\beta^{(k)}(v^{(k)}_l,\{v_i^{(k-1)}\})}
     \frac{sin(2v^{(k)}_l+d\eta)}{sin(\eta)} e^{i(d+k-2)\eta}\\[5mm]
 & &\cdot \displaystyle\frac{sin(2v^{(k)}_l+(k-1)\eta)}{sin(2v^{(k)}_l+k\eta)}
           \prod_{j=1,\neq l}^{P_k}\frac{sin(v^{(k)}_l+v^{(k)}_j+(k-1)\eta)}
           {sin(v^{(k)}_j-v^{(k)}_l-\eta)} \\[5mm]
                & &      ( 1\leq k \leq m)
\end{array}
\end{equation}

\begin{equation}
\begin{array}{rcl}
\Lambda^{(k+1)}(v_l^{(k)},\{v_i^{(k)}\},\{v_i^{(k+1)}\})
  &=&\displaystyle\frac{\alpha^{(k)}(v^{(k)}_l,\{v_i^{(k-1)}\})}{\beta^{(k)}(v^{(k)}_l,\{v_i^{(k-1)}\})}
       e^{i(d+2m-k)\eta}    \\[5mm]
   & & \displaystyle \frac{sin(2v^{(k)}_l+d\eta)sin(2v^{(k)}_l+(2m-k+1)\eta)}
       {sin(\eta)sin(2v^{(k)}_l+(2m-k)\eta)}
        \\[5mm] 
  & &\displaystyle \cdot\prod_{j=1,\neq l}^{P_k}\frac{sin(v^{(k)}_l+v^{(k)}_j+(2m-k+1)\eta)}
           {sin(v^{(k)}_l-v^{(k)}_j+\eta)} \\[2mm]
                & &      ( m+1\leq k \leq m+n-1)
\end{array}
\end{equation}
where
\begin{equation}
\begin{array}{rcl}
\alpha^{(k)}(u,\{v_i^{(k-1)}\})&=&\left\{\begin{array}{cc}
   \displaystyle \prod_{j=1}^{P_{k-1}}\frac{sin(u+v_j^{(k-1)}+k\eta)}{sin(v_j^{(k-1)}-u+\eta)},& 1\leq k\leq
                         m\\[5mm]
   \displaystyle\prod_{j=1}^{P_{k-1}}\frac{sin(u+v_j^{(k-1)}+(2m-k)\eta)}{sin(v_j^{(k-1)}-u-\eta)},& 
                   m+1\leq k\leq n  \end{array} \right.  \\[5mm]
\beta^{(k)}(u,\{v_i^{(k-1)}\})&=&\left\{\begin{array}{r}
                                 \displaystyle
                \prod_{j=1}^{P_{k-1}}\frac{sin(u+v_j^{(k-1)}+(k-1)\eta)sin(u-v_j^{(k-1)})}
                 {sin(v_j^{(k-1)}-u+\eta)sin(u-v_j^{(k-1)}+\eta)}\\[5mm]
                 \displaystyle\cdot\frac{sin(2u+(k-1))\eta)}{sin(2u+k\eta)}e^{-i\eta},
                  ~~1\leq k\leq m  \\[5mm]
                \displaystyle\prod_{j=1}^{P_{k-1}}\frac{sin(u+v_j^{(k-1)}+(2m-k+1)\eta)sin(u-v_j^{(k-1)})}
                 {sin(v_j^{(k-1)}-u+\eta)sin(u-v_j^{(k-1)}+\eta)}\\[5mm]
                 \displaystyle\cdot\frac{sin(2u+(2m-k+1))\eta)}{sin(2u+(2m-k)\eta)}
                    e^{-i\eta}, ~~m+1\leq k\leq n \end{array} \right. 
\end{array}
\end{equation}
In the above representation, we have assumed $v_j^{(1)}=v_j, v_j^{(0)}=0$, $P_o=N,P_1=L$ and
$\Lambda^{(m+n+1)}=1$. Notice that $\beta^{(k)}(u,\{v_i^{k-1}\})$ vanishes at the special points
$v_i^{(k-1)}$ due to the factor $sin(u-v_i^{(k-1)})$ appearing in $\beta^{(k)}$.
Taking $u=v_i^{(k-1)}$ in formulae (58) and (59), we can get another kind of constraints on
$\Lambda^{(k)}$
\begin{equation}
\begin{array}{rcl}
\Lambda^{(k)}(v_l^{(k-1)},\{v_i^{(k-1)}\},\{v_i^{(k)}\})
&=&\displaystyle\alpha^{(k)}(v_l^{(k-1)},\{v_i^{(k-1)}\})e^{i(d+k-2)\eta}
   \frac{sin(2v_l^{(k-1)}+d\eta)}{sin(2v_l^{(k-1)}+k\eta)}\\[5mm]
& &\displaystyle\prod_{j=1}^{P_k}\frac{sin(v_j^{(k)}-v_l^{(k-1)}+\eta)sin(v_j^{(k)}+v_l^{(k-1)}+(k-1)\eta)}
   {sin(v_j^{(k)}-v_l^{(k-1)})sin(v_j^{(k)}+v_l^{(k-1)}+k\eta)}\\[5mm]
& &1\leq k\leq m
\end{array}
\end{equation}

\begin{equation}
\begin{array}{rcl}
\Lambda^{(k)}(v_l^{(k-1)},\{v_i^{(k-1)}\},\{v_i^{(k)}\})
&=&-\displaystyle\alpha^{(k)}(v_l^{(k-1)},\{v_i^{(k-1)}\})e^{i(d+2m-k)\eta}
   \frac{sin(2v_l^{(k-1)}+d\eta)}{sin(2v_l^{(k-1)}+(2m-k)\eta)}\\[5mm]
& &\displaystyle\prod_{j=1}^{P_k}
  \frac{sin(v_j^{(k)}-v_l^{(k-1)}-\eta)sin(v_j^{(k)}+v_l^{(k-1)}+(2m-k+1)\eta)}
   {sin(v_j^{(k)}-v_l^{(k-1)})sin(v_j^{(k)}+v_l^{(k-1)}+(2m-k)\eta)}\\[5mm]
& &m+1\leq k\leq m+n
\end{array}
\end{equation}

Now, changing the index $k$ into $k+1$ in the above formulae, we can obtain constrains on
$\Lambda^{(k+1)}$. Comparing these with equations (60) and (61), one can derive out the following
Bethe ansatz equations
\begin{equation}
\begin{array}{l}
\displaystyle\prod_{j=1}^{P_{k-1}}
\frac{sin(v_l^{(k)}-v_j^{(k-1)}+\eta)sin(v_l^{(k)}+v_j^{(k-1)}+k\eta)}
     {sin(v_l^{(k)}-v_j^{(k-1)})sin(v_l^{(k)}+v_j^{(k-1)}+(k-1)\eta)}\\[5mm]
\cdot\displaystyle\prod_{j=1}^{P_{k+1}}
\frac{sin(v_l^{(k)}-v_j^{(k+1)}-\eta)sin(v_l^{(k)}+v_j^{(k+1)}+k\eta)}
     {sin(v_l^{(k)}-v_j^{(k+1)})sin(v_l^{(k)}+v_j^{(k+1)}+(k+1)\eta)}\\[5mm]
=\displaystyle\prod_{j=1,\neq l}^{P_k}
\frac{sin(v_l^{(k)}-v_j^{(k)}-\eta)sin(v_l^{(k)}+v_j^{(k)}+(k-1)\eta)}
     {sin(v_l^{(k)}-v_j^{(k)}+\eta)sin(v_l^{(k)}+v_j^{(k)}+(k+1)\eta)}\\[5mm]
~~~~~(1\leq k\leq m-1)
\end{array}
\end{equation}

\begin{equation}
\begin{array}{l}
\displaystyle\prod_{j=1}^{P_{m-1}}
\frac{sin(v_l^{(m)}-v_j^{(m-1)}+\eta)sin(v_l^{(m)}+v_j^{(m-1)}+m\eta)}
     {sin(v_l^{(m)}-v_j^{(m-1)})sin(v_l^{(m)}+v_j^{(m-1)}+(m-1)\eta)}\\[5mm]
~\cdot\displaystyle\prod_{j=1}^{P_{m+1}}
\frac{sin(v_l^{(m)}-v_j^{(m+1)})sin(v_l^{(m)}+v_j^{(m+1)}+(m-1)\eta)}
     {sin(v_l^{(m)}-v_j^{(m+1)}+\eta)sin(v_l^{(m)}+v_j^{(m+1)}+m\eta)}\\[5mm]
=1
\end{array}
\end{equation}

\begin{equation}
\begin{array}{l}
\displaystyle\prod_{j=1}^{P_{k-1}}
\frac{sin(v_l^{(k)}-v_j^{(k-1)}-\eta)sin(v_l^{(k)}+v_j^{(k-1)}+(2m-k)\eta)}
     {sin(v_l^{(k)}-v_j^{(k-1)})sin(v_l^{(k)}+v_j^{(k-1)}+(2m-k+1)\eta)}\\[5mm]
\cdot\displaystyle\prod_{j=1}^{P_{k+1}}
\frac{sin(v_l^{(k)}-v_j^{(k+1)})sin(v_l^{(k)}+v_j^{(k+1)}+(2m-k-1)\eta)}
     {sin(v_l^{(k)}-v_j^{(k+1)}+\eta)sin(v_l^{(k)}+v_j^{(k+1)}+(2m-k)\eta)}\\[5mm]
=\displaystyle\prod_{j=1,\neq l}^{P_k}
\frac{sin(v_l^{(k)}-v_j^{(k)}-\eta)sin(v_l^{(k)}+v_j^{(k)}+(2m-k-1)\eta)}
     {sin(v_l^{(k)}-v_j^{(k)}+\eta)sin(v_l^{(k)}+v_j^{(k)}+(2m-k+1)\eta)}\\[5mm]
~~~~~(m+1\leq k\leq m+n-1)
\end{array}
\end{equation}
The above Bethe ansatz equations are very complicted, but they can be
simplified by introducing the
following new variables
\begin{equation}
v_j^{(k)}=\left\{ \begin{array}{rc}
                 w_i^{(k)}-k\eta/2,& 1\leq k\leq m\\[5mm]
                 w_i^{(k)}-(2m-k)\eta/2,& m+1\leq k\leq m+n
                  \end{array} \right.
\end{equation}
The Bethe Ansatz equations then take the form

\begin{equation}
\begin{array}{l}
\displaystyle\prod_{j=1}^{P_{k-1}}
\frac{sin(v_l^{(k)}-v_j^{(k-1)}-\eta/2)sin(v_l^{(k)}+v_j^{(k-1)}-\eta/2)}
     { sin(v_l^{(k)}-v_j^{(k-1)}+/2\eta)sin(v_l^{(k)}+v_j^{(k-1)}+\eta/2)}\\[5mm]
\cdot\displaystyle\prod_{j=1}^{P_{k+1}}
\frac{sin(v_l^{(k)}-v_j^{(k+1)}-\eta/2)sin(v_l^{(k)}+v_j^{(k+1)}-\eta/2)}
     { sin(v_l^{(k)}-v_j^{(k+1)}+\eta/2)sin(v_l^{(k)}+v_j^{(k+1)}+\eta/2)}\\[5mm]
=\displaystyle\prod_{j=1,\neq l}^{P_k}
\frac{sin(v_l^{(k)}-v_j^{(k)}-\eta)sin(v_l^{(k)}+v_j^{(k)}-\eta)}
     { sin(v_l^{(k)}-v_j^{(k)}+\eta)sin(v_l^{(k)}+v_j^{(k)}+\eta)}\\[5mm]
~~~~~(1\leq k\leq m+n-1, k\neq m)
\end{array}
\end{equation}
\begin{equation}
\begin{array}{l}
\displaystyle\prod_{j=1}^{P_{m-1}}
\frac{sin(v_l^{(m)}-v_j^{(m-1)}-\eta/2)sin(v_l^{(m)}+v_j^{(m-1)}-\eta/2)}
     { sin(v_l^{(m)}-v_j^{(m-1)}+/2\eta)sin(v_l^{(m)}+v_j^{(m-1)}+\eta/2)}\\[5mm]
\cdot\displaystyle\prod_{j=1}^{P_{m+1}}
\frac{sin(v_l^{(m)}-v_j^{(m+1)}+\eta/2)sin(v_l^{(m)}+v_j^{(m+1)}+\eta/2)}
     { sin(v_l^{(m)}-v_j^{(m+1)}-\eta/2)sin(v_l^{(m)}+v_j^{(m+1)}-\eta/2)}\\[5mm]
=1
\end{array}
\end{equation}
The function $\Lambda^{(1)}(u,\cdots)$ must not be singular at $u=v_j^{(k)}$ ($1\leq j\leq p_k$,
$1\leq k\leq m+n-1$) since the transfer matrix $t(u)$ is an analytic function of $u$. In fact,
the equation (57) comes from the condition under which the unwanted term vanishes. One can
understand this constraint from another point of view: From equation (56), we know that
$u=v_j=v_j^{(1)}$ is a pole of $\Lambda^{(1)}(u)$. In order to keep the analyticity 
of $\Lambda^{(1)}(u)$, one should need the residuce of $\Lambda^{(1)}(u)$ at $v_j$ vanishing, which
also gives the constraint (57). So, $\Lambda^{(1)}(u)$ is analytic at $v_j$. Similarly, equations
(69) ensure the analyticity of $\Lambda^{(1)}(u)$ at all $v_j^{(k)}$. Therefore, the eigenvalues
of the transfer matrix are analytic functions if the previous Bethe ansatz equations are satisfied.

The energy spectrum of 1-dimensional quantum system defined by equation (23) can be derived fron
$\Lambda^{(1)}(u)$. It is 
\begin{equation}
E=-\sum_{k=1}^{P_1=L}\frac{1}{cos(\eta+2v_k)-cos(\eta)}+\frac{sin[(d-1)\eta]}{4sin^3(\eta)}
\end{equation}

In order to comparing our results with the Bethe Ansatz equations given in references [20,21], we
introduce new varibles
\begin{eqnarray*}
\lambda_j^k&=&\left\{\begin{array}{cc}
                      v^{(k)}_j~, & 1\leq j\leq P_k \\
                     -v^{(k)}_{2P_k-j+1}~,& P_k+1\leq j \leq 2P_k
                      \end{array} \right.
\end{eqnarray*}
Then equations (69) and (70) deduce to the Bethe Ansatz equations (see, for example, equation (4)
in ref. [20]) up to a phase.  In this sense, the Bethe Ansatz equations for the system with
quantum group symmetry is the duble of the ones for the same system with periodic boundary condition.
One should note the constrain in the right hand side of equation (69), which will contribute a 
non-zero term to the free energy. 

\section {Quantum group structure of the model}

In this section we will show that the vertex model under consideration
is a realization of quantum supergroup $SU_q(n|m)$, and we will also prove that the transfer matrix 
for open boundary conditions is $SU_q(n|m)$ invariant.

Firstly, denoting $x=e^{iv}$, $q=e^{i\eta }$, the Yang-Baxter equation 
becomes

\begin{equation}
R_{12}(x/y)T_1(x)T_2(y)=T_2(y)T_1(x)R_{12}(x/y).
\end{equation}

\noindent We write the R-matrix as

\begin{eqnarray}
R(x)=xR_+-x^{-1}R_-,
\end{eqnarray}

\noindent similaryly, the $L$ operators can be written as 

\begin{eqnarray}
L(x)=xL_+-x^{-1}L_-.
\end{eqnarray}

\noindent From the definition of R-matrix, $L_{\pm }$ can be written in 
the following form.

\begin{eqnarray}
{L_+}^i_i=\left\{ \begin{array}{ll}
q^{w_i}, &i\leq m,\\
\sigma _iq^{-w_i}, &m<i\leq m+n,\end{array}
\right.
\end{eqnarray}

\begin{eqnarray}
{L_+}^{i+1}_i=\left\{ \begin{array}{ll}
(q-q^{-1})q^{-{1\over 2}\sum_{j\not= i,i+1}w_j}f_i, &i<m,\\
(q-q^{-1})q^{-{1\over 2}\sum_{j\not= m,m+1}w_j-w_{m+1}}\sigma _mf_m, &i=m,\\
(q-q^{-1})q^{{1\over 2}\sum_{j\not= i,i+1}w_j}\sigma _if_i, &m<i\leq m+n,
\end{array}\right.
\end{eqnarray}

\begin{eqnarray}
{L_-}^i_i=\left\{ \begin{array}{ll}
q^{-w_i}, &i\leq m,\\
\sigma _iq^{w_i}, &i>m,
\end{array} \right.
\end{eqnarray}

\begin{eqnarray}
{L_-}^i_{i+1}=\left\{ \begin{array}{ll}
-(q-q^{-1})e_iq^{{1\over 2}\sum _{j\not= i,i+1}w_j}, &i<m,\\
-(q-q^{-1})e_mq^{{1\over 2}\sum _{j\not= m,m+1}w_j+w_{m+1}}, &i=m\\
-(q-q^{-1})\sigma _ie_iq^{-{1\over 2}\sum _{j\not= i,i+1}w_j}, &i>m.
\end{array} \right.
\end{eqnarray}

\noindent Here $L_{\pm }$ are lower and upper triangular matrices with 
${L_+}^i_j={L_-}^j_i=0$, if $i<j$, $w_i$,$i=1,\cdots ,m+n$;
$e_i, f_i$, $i=1,\cdots ,m+n-1$, are the generators of the $SU(n|m)$ 
superalgebra in the graded Cartan-Chevalley basis; the definition of 
the matrices $\sigma _i$ and the details of the classical simple Lie algebra
$SU(n|m)$ are given in Appendix B. Recall the definition of the monodromy
matrix $T(x)$: In the limit $x\rightarrow \infty ,0$, we find the 
leading terms $T_{\pm }$ of the monodromy matrix $T(x)$ to take the form,

\begin{eqnarray}
{T_+}^i_i=\left\{ \begin{array}{ll}
q^{-N/2}q^{w_i}, &i\leq m,\\
q^{-N/2}\sigma _iq^{-W_i}, &m<i\leq m+n,
\end{array}\right.
\end{eqnarray}

\begin{eqnarray}
{T_+}^{i+1}_i=\left\{ \begin{array}{ll}
\alpha _-q^{-{1\over 2}\sum_{j\not= i,i+1}W_j}F_i, &i<m,\\
\alpha _-q^{-{1\over 2}\sum_{j\not= m,m+1}W_j-W_{m+1}}\tilde{\sigma }_mF_m,
&i=m,\\
q^{-N}\alpha _+q^{{1\over 2}\sum _{j\not=i,i+1}W_j}\tilde{\sigma }_i
F_i, &i>m,
\end{array}\right.
\end{eqnarray}

\begin{eqnarray}
{T_-}^i_i=\left\{ \begin{array}{ll}
q^{N/2}q^{-W_i}, &i\leq m,\\
q^{N/2}\sigma _iq^{W_i}, &m<i\leq m+n,
\end{array}\right.
\end{eqnarray}

\begin{eqnarray}
{T_-}^i_{i+1}=\left\{ \begin{array}{ll}
-\alpha _+E_iq^{{1\over 2}\sum_{j\not= i,i+1}W_j}, &i\leq m,\\
-\alpha _+E_mq^{{1\over 2}\sum_{j\not= m,m+1}W_j+W_{m+1}}, &i=m,\\
-q^N\alpha _-\tilde{\sigma }_iE_iq^{-{1\over 2}\sum_{j\not =i,i+1}W_j}, 
&m<i\leq m+n.
\end{array}\right.
\end{eqnarray}

\noindent Here $T_{\pm }$ are lower and upper triangular matrices with
${T_+}^i_j={T_-}^j_i=0$, if $i<j$, $\alpha _{\pm }=q^{\pm 1/2}(q-q_{-1})$, and

\begin{eqnarray}
\tilde{\sigma }_i&=&\sigma _i\otimes \sigma _i\otimes \cdots \otimes 
\sigma _i, i=m+1,\cdots ,m+n,\nonumber \\
q^{\pm W^i}&=&q^{\pm w_i}\otimes \cdots \otimes q^{\pm w_i}, 
i=1,\cdots ,m+n, \nonumber \\
X_i&=&\sum_{j=1}^{N}q^{-h_i/2}\otimes \cdots \otimes q^{-h_i/2}
\otimes x_i^{j_{th}}\otimes q^{h_i/2}\otimes \cdots \otimes q^{h_i/2}, 
i<m, \\
X_m&=&\sum_{j=1}^{N}q^{-h_m/2}\otimes \cdots \otimes q^{-h_m/2}
\otimes x_m^{j_{th}}\otimes (\sigma _{m+1}q^{h_m/2})
\otimes \cdots \otimes (\sigma _{m+1}q^{h_m/2}),\nonumber \\
X_i&=&\sum_{j=1}^Nq^{h_i/2}\otimes \cdots \otimes q^{h_i/2}
\otimes x_i^{j_{th}}\otimes (\sigma _{m+1}\sigma _{m+2}\cdots 
\sigma _{i+1}q^{-h_i/2})\nonumber \\
& &\otimes \cdots \otimes (\sigma _{m+1}
\sigma _{m+2}\cdots \sigma _{i+1}q^{-h_i/2}), m<i\leq m+n.\nonumber
\end{eqnarray}

\noindent where $X_i=E_i,F_i$, $x_i=e_i,f_i$ , respectively. 
$h_i=w_i-w_{i+1}$, $i\not= m$, and $h_m=w_m+w_{m+1}$. In the case of 
$N=2$, these formulae define the coproduct of a Hopf algebra.
From this point of view, the equation (83) can be written as  

\begin{eqnarray}
X_i=\Delta ^{N-1}(x_i)=(\Delta \otimes id)\Delta ^{N-2}(x_i)
\end{eqnarray}

In the following we will discuss the algebraic relations of ($q^{w_i}, X_i$). Taking
the appropriate limits of the  R-matrix and  the row-to-row monodromy matrix $T$, we have

\begin{eqnarray}
\lim_{x \to 0}xR(x/y)&=&-yR_-\\
\lim_{x \to \infty }x^{-1}R(x/y)&=&{1\over y}R_+
\end{eqnarray}

\noindent and 

\begin{eqnarray}
\lim_{x \to }x^NT(x)&=&-T_-\\
\lim_{x \to \infty }x^{-N}T(x)&=&T_+
\end{eqnarray}

\noindent In the limits $x\rightarrow 0,\infty $,
the Yang-Baxter equation gives:

\begin{equation}
R_{\pm }T_{1\pm }T_2(y)=T_2(y)T_{1\pm }R_{\pm }
\end{equation}

\noindent and

\begin{equation}
R_{\pm }T_{1\pm }T_{2\varepsilon }=T_{2\varepsilon }T_{1\pm }R_{\pm } 
\end{equation}

\noindent with $\varepsilon =\{+,-\}$. These  
spectral-parameter-indepedent Yang-Baxter relations govern 
q-(anti)commutation rules and q-Serre relations for the quantum supergroup
$SU_q(n|m)$. 
Substituting the definition of $R_{\pm},T_{\pm}$ into equation (90), we get
\begin{eqnarray}
q^{H_i}q^{H_j}&=&q^{H_j}q^{H_i},\nonumber \\
q^{H_i}F_jq^{-H_i}&=&q^{a_{ij}}F_j,\nonumber \\
q^{H_i}E_jq^{-H_i}&=&q^{-a_{ij}}E_j,\nonumber \\
\left [F_i,E_i\right ]&=&\frac {q^{H_i}-q^{-H_i}}{q-q^{-1}}, i\ne m,\nonumber \\
\left[ F_m,E_m\right] _+&=&\frac {q^{H_m}-q^{-H_m}}{q-q^{-1}}\nonumber \\
E_m^2=F_m^2=0, [F_i,E_j]&=&0, i\ne j, \nonumber \\
(F_i)^2F_{i\pm 1}-(q+q^{-1})F_iF_{i\pm 1}F_i+F_{i\pm 1}(F_i)^2&=&0,\nonumber \\
(E_i)^2E_{i\pm 1}-(q-q^{-1})E_iE_{i\pm 1}E_i+E_{i\pm 1}(E_i)^2&=&0.\nonumber \\
\end{eqnarray}

\noindent where $H_i=W_i-W_{i+1}$, if $i\ne m$, $H_m=W_m+W_{m+1}$, and $a_{ij}$
is a component of the Cartan matrix which is given in Appendix B. The
generators $H_i,E_i, F_i$, $i=1,\cdots ,m+n-1$, and relations listed
above provide a definition of the quantum supergroup $SU_q(n|m)$. In the remaining part  
of this section, we will verify that the transfer matrix $t(y)$ with open
boundary conditions is $SU_q(n|m)$ invariant. The 
entries of lower and uper triangular matrix $T_{\pm }$ are elements
of $SU_q(n|m)$. So, it is not necessary to compute commutators of 
$t(y)$ with individual $SU_q(n|m)$ generators. 
If the relation 
\begin{equation}
[t(y), T_{\pm }]=0
\end{equation}
\noindent is correct, we are led to the conclusion that the transfer matrix
$t(u)$ is $SU_q(n|m)$ invariant. From eq. (89) we have the result

\begin{eqnarray}
[R_{\pm }T_{1\pm }, T_2(y)T^{-1}_2(y^{-1})]=0
\end{eqnarray}

\noindent Recall the relation (12), we have

\begin{equation}
[R_{\pm }, M_1M_2]=0
\end{equation}

\noindent Similary, from the unitarity and cross-unitarity relations (9,10),
with the help of $PT$ invariance of R-matrix, we find

\begin{eqnarray}
R_{\pm }R^{t_1t_2}_{\mp }&=&1\nonumber \\
R^{t_1}_{\pm }M_1R_{\mp }^{t_2}M_1^{-1}&=&1
\end{eqnarray}
  
\noindent So we have the identity $R^{t_1}_{\mp }=(R_{\pm }^{-1})^{t_2}$ 
which implies that the following relation
\begin{equation}
M^{-1}_1(R^{-1}_{\pm })^{t_2}M_1R_{\pm }^{t_2}=1
\end{equation}
\noindent is correct. Notice that we choose $K_-=1$ in this paper, so the
transfer matrix can be written as $t(y)=trMT(y)T^{-1}(y^{-1})$. Now, let
us prove relation (92)

\begin{eqnarray}
T_{1\pm }t(y)&=&tr_2T_{1\pm }M_2T_2(y)T^{-1}_2(y^{-1})\nonumber \\
&=&tr_2M_2R_{\pm }^{-1}R_{\pm }T_{1\pm }T_2(y)T^{-1}_2(y^{-1})
\end{eqnarray}

\noindent here we have added an identity $R_{\pm }^{-1}R_{\pm }$ in the  
relation, then using the relations (95) and (96), we find

\begin{eqnarray}
\cdots&=&tr_2M_1^{-1}R_{\pm }^{-1}M_1(M_2T_2(y)T_2^{-1}(y^{-1}))R_{\pm }
T_{1\pm }\nonumber \\
&=&tr_2\{ M_1^{-1}R^{-1}_{\pm }M_1\} ^{t_2}\{ (M_2T_2(y)
T^{-1}_2(y^{-1}))R_{\pm }T_{1\pm }\} ^{t_2}\nonumber \\
&=&tr_2M_2T_2(y)T^{-1}_2(y^{-1})T_{1\pm } \nonumber \\
&=&t(y)T_{1\pm }.
\end{eqnarray}

 Thus, we have proved that the transfer 
matrix with a particular choice of open boundary conditions is 
quantum supergroup $SU_q(n|m)$ invariant.

\section{Summary}
 In this paper, we have diagonalized the graded vertex model with open boundary condition
by using the generalized algebraic Bethe ansatz method. In order to get the energy spectrum of
1-dimensional quantum system defined by equation (23), we assume $v_j^{(0)}$ to be zero in
equations (58) and (60). However, one would as will assume $v_j^{(0)}\neq 0$. In this case, 
equations (56), (69) and (70) lead to  the solution of inhomogeneous graded vertex model. 
Formally, one can also define a 1-dimensional quantum 
system by equation (23). Generally, the hamiltonian
is not represented in the nearest neighbour interaction form. We also show the $SU_q(m|n)$
invariance of the quantum spin chain (equivalent to a graded vertex model). Thus, the 
generators of $SU_q(m|n)$ commute with the infinite number of conserved quantityies. The Hilbert space
of the system can be classified according to the irreducible representations of $SU_q(m|n)$.
We hope that it will be help to solve the Bethe ansatz equations.

In order to find the free energy of the system, one should to solve the Bethe ansatz equations.
Following the method given in reference [18], we can deduce the Bethe ansatz equations 
into those of the periodic case on $2N$ sites with an additional source factor (see ref. [20]).
The free energy contains two terms. One is the known bulk free energy, another is the surface
free energy which is the correction of the open boundary conditions (that keeps the quantum
group symmetry). This was pointed by de Vega and Gonzalez-Ruiz in $SU(n)$ case.

\vspace{1cm} 
\noindent {\bf Acknowlegements}

R. Yue was supported by Alexander von Humboldt Fundation. 
He would like to thank Prof. Werner Nahm for the hospitality and encouragement.
We also thank Prof. K.Shi and Dr. Z.Yang for useful discussions.

\section{Appendix A}
The starting point for commutation relations is reflection equation (25).
Let $a_1=a_2=b_2=1$, $b_1=b\not= 1$, we find:

\begin{eqnarray}
A(v)B_b(u)&=&\frac {R_{12}(u_-)^{11}_{11}R_{21}(u_+)^{b1}_{b1}}
{R_{12}(u_+)^{11}_{11}R_{21}(u_-)^{b1}_{b1}}B_b(u)A(v)\nonumber \\
& &-\frac {R_{12}(u_+)^{1b}_{1b}R_{21}(u_-)^{1b}_{b1}}
{R_{12}(u_+)^{11}_{11}R_{21}(u_-)^{b1}_{b1}}B_b(v)A(u)\nonumber \\
& &-\frac {R_{12}(u_+)^{1c}_{c1}}{R_{12}(u_+)^{11}_{11}}B_c(v)D_{cb}(u)
\end{eqnarray}

\nonumber Due to eq.(27), it can be checked that the following relation
is always true for $R(u)^{11}_{11}=sin(\eta \pm u)$.

\begin{eqnarray}
A(v)B_b(u)&=&\frac {R_{12}(u_-)^{11}_{11}R_{21}(u_+)^{b1}_{b1}}
{R_{12}(u_+)^{11}_{11}R_{21}(u_-)^{b1}_{b1}}B_b(u)A(v)\nonumber \\
& &-\frac {R_{12}(u_-)^{b1}_{1b}R_{12}(2u)^{1b}_{1b}}
{R_{12}(2u)^{11}_{11}R_{12}(u_-)^{1b}_{1b}}B_b(v)A(u)\nonumber \\
& &-\frac {R_{12}(u_+)^{1c}_{c1}}{R_{12}(u_+)^{11}_{11}}B_c(v)
\tilde{D}_{cb}(u)
\end{eqnarray}

\noindent Obviously, commutation relations (29,31) can be obtained from (100).

Next, let $a_2=1$, $a_1,b_1,b_2\not= 1$, one can get:

\begin{eqnarray}
D_{a_1b_1}(u)B_{b_2}(v)&=&\frac {R_{12}(u_+)^{a_1c_2}_{c_1d_2}
R_{21}(u_-)^{d_1d_2}_{b_1b_2}}{R_{12}(u_-)^{a_11}_{a_11}
R_{21}(u_+)^{b_11}_{b_11}}B_{c_2}(v)D_{c_1d_1}(u)\nonumber \\
& &-\frac {R_{12}(u_-)^{a_11}_{1a_1}R_{21}(u_+)^{d_1a_1}_{b_1d_2}}
{R_{12}(u_-)^{a_11}_{a_11}R_{21}(u_+)^{b_11}_{b_11}}
B_{d_1}(u)D_{d_2b_2}(v).
\end{eqnarray}

\noindent Substituting (27) to (A.3), we obtain

\begin{eqnarray}
\tilde{D}_{a_1b_1}(u)B_{b_2}(v)&=&{1\over {R_{12}(u_-)^{a_11}_{a_11}
R_{21}(u_+)^{b_11}_{b_11}}}\{ R_{12}(u_+)^{a_1c_2}_{c_1d_2}
R_{21}(u_-)^{d_1d_2}_{b_1b_2}B_{c_2}(v)\tilde{D}_{c_1d_1}(u)\nonumber \\
& &-R_{12}(u_-)^{a_11}_{1a_1}R_{21}(u_+)^{d_1a_1}_{b_1d_2}B_{d_1}(u)
\tilde{D}_{d_2b_2}(v)\} \nonumber \\
& &+{1\over {R_{12}(u_-)^{a_11}_{a_11}R_{21}(u_+)^{b_11}_{b_11}}}F
\end{eqnarray}

\begin{eqnarray}
F&=&R_{12}(u_+)^{a_1c_2}_{c_1d_2}R_{21}(u_-)^{c_1d_2}_{b_1b_2}
{\frac {R_{12}(2u)^{c_11}_{1c_1}}{R_{12}(2u)^{11}_{11}}}
B_{c_2}(v)A(u)\nonumber \\
& &-R_{12}(u_-)^{a_11}_{1a_1}R_{21}(u_+)^{d_1a_1}_{b_1b_2}
{\frac {R_{12}(2v)^{b_21}_{1b_2}}{R_{12}(2v)^{11}_{11}}}
B_{d_1}(u)A(v)\nonumber \\
& &-(R_{12}(u_-)^{a_11}_{a_11}R_{21}(u_+)^{b_11}_{b_11}
{\frac {R_{12}(2u)^{a_11}_{1a_1}}{R_{12}(2u)^{11}_{11}}}\nonumber \\
& &+R_{12}(u_-)^{a_11}_{1a_1}R_{21}(u_+)^{1a_1}_{a_11})
\delta _{a_1b_1}A(u)B_{b_2}(v)\nonumber \\
& &+R_{12}(u_+)^{a_11}_{1a_1}R_{21}(u_-)^{d_1a_1}_{b_1b_2}A(v)B_{d_1}(u)
\end{eqnarray}

\noindent In the following we will calculate the function $F$ for the case of 
$R(u)^{11}_{11}=a(u)=sin(u+\eta )$. The results can be written in a 
simple form, the main calculation results are as follows.

\noindent Case 1: $a_1\not= b_1$

\begin{eqnarray}
F&=&\delta _{a_1b_2}{\frac {sin\eta e^{i(u+v)}sin(2v)sin(u+v)sin(u-v)}
{sin(u+v+\eta )sin(2v+\eta )}}B_{b_1}(u)A(v)\nonumber \\
& &-\delta _{a_1b_2}{\frac {sin^2(\eta )sin(u-v)}{sin(u+v+\eta )}}
B_c(v)\tilde D_{cb_1}(u)
\end{eqnarray}

\noindent Case 2: $a_1=b_1=b_2=a$

\begin{eqnarray}
F&=&{\frac {sin(\eta )e^{i(u+v)}sin(u+v)sin(u-v)sin(2v)sin(\eta +
\epsilon _a(2u+\eta ))}{sin(2u+\eta )sin(2v+\eta )sin(u+v+\eta )}}
B_a(u)A(v)\nonumber \\
& &-{\frac {sin^2(\eta )sin(\eta +\epsilon _a(u-v))}{sin(u+v+\eta )}}
B_c(v)\tilde{D}_{ca}(u)\nonumber \\
& &+{\frac {sin(u-v+\eta )sin^2(\eta )e^{i(u-v)}}
{sin(2u+\eta )}}B_c(u)\tilde{D}_{ca}(v)
\end{eqnarray}

\noindent Case 3: $a_1=b_1\not= b_2$

\begin{eqnarray}
F&=&{\frac {sin(\eta )e^{i(u+v)}sin(u+v)sin(u-v)sin(2v)
R(2u+\eta )^{a_1b_2}_{b_2a_1}}{sin(2v+\eta )sin(2u+\eta )sin(u+v+\eta )}}
B_{b_2}(u)A(v)\nonumber \\
& &-{\frac {sin^2(\eta )R_{21}(u-v)^{b_2a_1}_{a_1b_2}}
{sin(u+v+\eta )}}B_c(v)\tilde{D}_{cb_2}(u)\nonumber \\
& &+{\frac {sin(u-v+\eta )sin^2(\eta )e^{i(u-v)}}
{sin(2u+\eta )}}B_c(u)\tilde{D}_{cb_2}(v)
\end{eqnarray}

\noindent Correspondingly, in the case of 
$R(u)^{aa}_{aa}=sin(\eta -u)=w(u), a=1,\cdots ,n$, the main calculation
results are presented in the following form.

\noindent Case 1: $a_1\not= b_1$

\begin{eqnarray}
F&=&\delta _{a_1b_2}{\frac {sin(\eta )e^{i(u+v)}sin(2v)sin(v+v)sin(u-v)}
{sin(\eta -u-v)sin(\eta -2v)}}B_{b_1}(u)A(v)\nonumber \\
& &-\delta _{a_1b_2}{\frac {sin^2(\eta )sin(u-v)}{sin(\eta -u-v)}}B_c(v)
\tilde{D}_{cb_1}(u)
\end{eqnarray}

\noindent Case 2: $a_1=b_1=b_2=a$

\begin{eqnarray}
F&=&{\frac {sin(\eta )e^{i(u+v)}sin(u+v)sin(2v)sin(2u-2\eta )sin(u-v)}
{sin(\eta -2v)sin(\eta -2u)sin(\eta -u-v)}}B_a(u)A(v)\nonumber \\
& &-{\frac {sin^2(\eta )sin(\eta -u+v)}{sin(\eta -u-v}}
B_c(v)\tilde{D}_{ca}(u)\nonumber \\
& &+{\frac {sin^2(\eta )e^{i(u-v)}sin(\eta -u+v)}
{sin(\eta -2u)}}B_c(u)\tilde{D}_{ca}(v)
\end{eqnarray}

\noindent Case 3: $a_1=b_1\not =b_2$

\begin{eqnarray}
F&=&{\frac {sin(u+v)sin(u-v)sin(2v)sin(\eta )e^{i(u+v)}
R_{12}(2u-\eta )^{a_1b_2}_{b_2a_1}}{sin(2u-\eta )sin(\eta -2v)
sin(\eta -u-v)}}B_{b_2}(u)A(v)\nonumber \\
& &-{\frac {sin^2(\eta )R_{21}(u-v)^{b_2a_1}_{a_1b_2}}
{sin(\eta -u-v)}}B_c(v)\tilde{D}_{cb_2}(u)\nonumber \\
& &+{\frac {sin(\eta -u+v)sin^2(\eta )e^{i(u-v)}}
{sin(2u-\eta )}}B_c(u)\tilde{D}_{cb_2}(v)
\end{eqnarray}

Though we have already simplified the results, they still
seem to be too complicated to be dealt with. Fortunately,
we have found that the results (104-109) can be summarized as a concise form
which indicates the commutation rules between $\tilde{D}_{a_1b_1}(u)$
and $B_{b_2}(v)$. The explicit commutation relations are written in sect.3.
One can prove it by expanding relations (30,32) according to different
cases mentioned above. Thus, we have obtained the commutation relations
(29-32). It is also necessary to calculate the commutaion relations 
between $B_a(u)$ and $B_b(v)$.

Let $a_1=a_2=1$, $b_1, b_2\ne 1$, we have the results:

\begin{eqnarray}
B_{b_1}(u)B_{b_2}(v)=\frac {R_{12}(u_+)^{1c_2}_{1c_2}
R_{21}(u_-)^{d_1c_2}_{b_1b_2}}{R_{12}(u_-)^{11}_{11}
R_{21}(u_+)^{b_11}_{b_11}}B_{c_2}(v)B_{d_1}(u)
\end{eqnarray}

\section{Appendix B}

The classical simple graded Lie algebra $SU(n|m)$ is defined by 
generators $h_i,e_i,f_i, i=1,\cdots, m+n-1$ and the following
relations

\begin{eqnarray}
\left [h_i,h_j\right ]&=&0,\nonumber \\
\left [h_i,f_j\right ]&=&a_{ij}f_j, [h_i,e_j]=-a_{ij}e_j,\nonumber \\
\left [f_i,e_i\right ]&=&h_i,i\ne m,\nonumber \\
\left [f_m,e_m\right ]_+&=&h_m,\nonumber \\
\left [f_i,e_j\right ]&=&0,i\ne j,\nonumber \\
f_m^2=e_m^2&=&0,\nonumber \\
f_i^2f_{i\pm 1}-2f_if_{i\pm 1}f_i+f_{i\pm 1}f_i^2&=&0\nonumber \\
e_i^2e_{i\pm 1}-2e_ie_{i\pm 1}e_i+e_{i\pm 1}e_i^2&=&0
\end{eqnarray}

\noindent The last two relations are the so called Serre relations,  
which are  compatible conditions for $SU(n|m)$, $a_{ij}$ is the
component of the graded Cartan matrix A  defined by:

\begin{eqnarray}
a_{ii}&=&2,i\ne m,\nonumber \\
a_{mm}&=&0,\nonumber \\
a_{i+1,i}&=&-1\nonumber \\
a_{i,i+1}&=&-1, i\ne m \nonumber \\
a_{m,m+1}&=&1
\end{eqnarray}

\noindent the other elements being equal to zero. We define $\sigma _i$
as:

\begin{equation}
\sigma _i=diag(1,1,\cdots ,1,-1,1,\cdots ,1)
\end{equation}
where -1 is th $i$th element.

The fundamental representation of the generators takes the form

\begin{eqnarray}
w_i&=&E_{i,i},i=1,\cdots ,m+n,\nonumber \\
f_i&=&E_{i,i+1}, i=1,\cdots ,m+n-1,\nonumber \\
e_i&=&E_{i+1,i}, i=1, \cdots ,m+n-1, \nonumber \\
h_i&=&w_i-w_{i+1}, i\ne m,\nonumber \\
h_m&=&w_m+w_{m+1}.
\end{eqnarray}

\noindent Here $E_{ij}$ are  $(m+n)\times (m+n)$ matrices with the 
element in $i$-row $j$-column equal to 1, all other elements being zero.


\begin{thebibliography}{25}
\bibitem{1}R.J.Baxter, Exactly Solved Models in Statistical Mechanics, 
Academic Press, New York.1982.
\bibitem{2}C.N.Yang, Phys.Rev.Lett.19(1967)1312. 
\bibitem{3}I.V.Cherednik, Theor.Math.Phys. 17(1983)77;61(1984)911.
\bibitem{4}E.K.Sklyanin, J.Phys.A21(1988)2375. 
\bibitem{5}C.Destri, H.J.de Vega, Nucl.Phys.B361(1992);B374(1992)692.
\bibitem{6}P.P.Kulish, E.K.Sklyanin, J.Phys.A24(1991)L435-L439.
\bibitem{7}L.Mezincescue, R.I.Nepomechie, J.Phy.A24(1991)L19;\\
Int.J.Mod.Phys.Lett.A6(1991)2497.
\bibitem{8}V.V.Bazhanov,Phys.Lett.B159(1985)321;
Commun.Math.Phys.113(1987)471
\bibitem{9}M.Jimbo,Commun.Math.Phys.102(1986)537.
\bibitem{10}H.Fan, B.Y.Hou, K.J.Shi, Z.X.Yang, to appear in Phys.Lett.A.
\bibitem{11}R.H.Yue, Y.X.Chen J. Phys. A26(1993)2989;
\bibitem{12}S.Ghoshel, A.B.Zamolodchikov, Int.J.Mod.Phys.A9(1994)3841;
\bibitem{13}E.Corrigan, P.E.Dorey, R.H.Rietdijk, R.Sasaki, Phys.Lett.B333(1994)83;
\bibitem{14}S.Ghoshel, Phys.Lett.B334(1994)363;
\bibitem{15}E.K.Sklyanin, Funct.Anal.Appl.21(1987)164;
\bibitem{16}A.Foerster, M.Karowski, Nucl.Phys.B396(1993)611;Nucl.Phys.B408(1993)[FS]512.
\bibitem{17}A.Gonzalez-Ruiz, Nucl.Phys.B424(1994)[FS]468.
\bibitem{18}H.J.de Vega, A.Gonzalez-Ruiz,Nucl.Phys.B417(1994)553.
\bibitem{19}J.H.Perk, C.L.Schultz. Phys.Lett.A84(1981)407.
\bibitem{20}H.J.de Vega, E.Lopes, Phys.Rev.Lett.67(1991)489.
\bibitem{21}E.Lopes, Nucl.Phys.B370(1992)636.
\bibitem{22}O.Babelon, H.J.de Vega, C.M.Viallet, Nucl.Phys.B200(1982)[FS]266.
\bibitem{23}F.H.Essler, V.E.Korepin, K.Schoutens, Phys.Rev.Lett.68(1992)2960.
\bibitem{24}H.J.de Vega, Int.J.Mod.Phys.A4(1989)2317.
\end{thebibliography}
\end{document}